\documentclass[prd,twocolumn,aps,amsmath,amssymb,showpacs,10pt]{revtex4-1}

\usepackage{amsbsy}
\usepackage{graphicx}
\usepackage{bm}
\usepackage{graphics}
\usepackage{latexsym}

\usepackage[active]{srcltx}

\newcommand{\sect}[1]{\setcounter{equation}{0}\section{#1}}

\renewcommand{\theequation}{\arabic{equation}}
\def\spazio#1{\vrule height#1em width0em depth#1em}

\def\spazio#1{\vrule height#1em width0em depth#1em}

\def\ta{\Bigl(\Bigr.}
\def\tc{\Bigl.\Bigr)}

\begin{document}

\begin{abstract}
{ A system of two fermions  with different masses and interacting by the Coulomb potential is presented in a
completely covariant framework. The spin-spin interaction, including the anomalous magnetic moments of the two fermions,
is added by means of a Breit term. We solve the resulting fourth order differential system by evaluating
the spectrum and the eigenfunctions. The interaction vertex with an external electromagnetic field is then
determined. The relativistic eigenfunctions are used to study the photon emission from a hyperfine transition and are
checked for the calculation of the Lamb shift due to the electron vacuum polarization in the muonic Hydrogen.}
\end{abstract}

\bigskip
\bigskip

\title{\bf Relativistic two fermion treatment of hyperfine transitions.}

\author{A. Barducci}
\affiliation{Dipartimento di Fisica, Universit\`a di Firenze,
Italy} \affiliation{Istituto Nazionale di Fisica Nucleare, Sezione
di Firenze} 
\author{R. Giachetti}
\affiliation{Dipartimento di Fisica, Universit\`a di Firenze,
Italy} \affiliation{Istituto Nazionale di Fisica Nucleare, Sezione
di Firenze} 
\author{E. Sorace}
\affiliation{Istituto Nazionale di Fisica Nucleare, Sezione di
Firenze}

\maketitle

\bigskip
\bigskip

%
%
%
%

%
\sect{Introduction} \label{Sec_introduction}
%

The precision physics of simple atomic systems has become a highly developed discipline both on the experimental and on
the theoretical side. Many new papers appear in each issue of the most important physics journals and a simple search on
the web gives millions of records. The accuracy by which theoretical calculations reproduce the
measured quantities is really remarkable and many effects of different types are being constantly added to improve the
agreement with experiments and to give a better and better determination of some fundamental physical constants. This,
of course, is very demanding on the amount of work to be done, as the approach generally used is a perturbation
expansion in the fine structure constant, starting with a non relativistic description of the atomic components: taking
into account the relativistic corrections, the presence of two different components of finite mass, the radiative terms
of quantum electrodynamics and the possible nuclear effects often requires the development of rather complex
analytical and numerical techniques (see, \emph{e.g.}, the book \cite{Eides} and the review paper \cite{Karshen}). Thus
our purpose is \emph{not} to compete on the accuracy and this paper is \emph{not} on precision atomic physics, al least
as it is commonly understood. We are more interested in posing the problem so as to incorporate completely the
fundamental properties of the physical system, namely the fact that we are dealing with a real two body system, that we
want to consider it in a completely covariant way and that the two component particles are fermions. 
The initial description \cite{GS1,GS2} dates some years back and was obtained in terms of two coupled Dirac equations.
The covariance was proved and it was shown that the resulting equation satisfies the correct Schr\"odinger limit for
infinite $c$. It also reproduces the one particle Dirac equation when one of the two masses tends to infinity,
overcoming some of the difficulties of the Bethe-Salpeter approach (see, eg. \cite{Lucha}). 
An evident drawback is that some systems, such as the deuterium, are excluded by our treatment which, however, includes
most of the interesting simple atoms. 
More recently the numerical aspects of our scheme were reconsidered and improved, especially in connection with the
application of the two fermion equation to the study of meson spectra by means of the Cornell potential and of a Breit
term, \cite{GS3}. The results have shown that a completely covariant description can produce a unified framework for the
case of light mesons too, where non relativistic approaches had failed; it was also made clear, contrary to common
wisdom, that a complete covariance was highly desirable in order to deal with mesons composed of quarks with different
masses \cite{Richard}, even when a non relativistic approximation seemed to have a good chance of being successful. 

Just as the Dirac equation for a single fermion in a central potential is written in terms of a pair of first
order differential equations, the two interacting fermions unavoidably give rise to a differential system of 
order four. This can cause some annoyance, not only because the most immediate insight into physical problems has
been developed in terms of second order equations, but also since most of the analytic and numeric
approximations have been mainly adapted to the Schr\"odinger equation. At present we have not found any really
sensible analytical method to discuss our fourth order system, but a completely numerical solution does not present
serious difficulties apart from the necessary care needed to meet the precision required by the calculations. The
benefits we get are rewarding: firstly, the complete and built-in inclusion of the relativistic features
such as the fine structure and a spin-orbit contribution for each component implied by each of
the two Dirac equations; secondly, the correct treatment of the finite mass for each of the two particles, thus
comprising the recoil effects for the pure Coulomb problem and avoiding the corrections due to the reduced mass. A
numerical non perturbative approach was recently proposed in \cite{Rafelski} where the authors used a single Dirac
equation with a reduced mass warning, however,  that the separation of the center of mass may cause some inconsistency
at the relativistic level. We finally introduce a Breit term in order to describe the spin-spin interaction responsible
for the hyperfine structure: we notice the difference with respect to \cite{GS2} because of the inclusion of the fermion
anomalous magnetic moments.   

The most relevant aspects this paper addresses to are twofold. We first present very concisely some numerical results
that we believe can be of some interest, being completely clear what they take exactly into account.
They are non perturbative in the Coulomb field and the hyperfine splittings (HFS) are given by the first
perturbation order in the Breit interaction. They are not hard to calculate and their accuracy, although incomparable
with the results existing in literature, is however rather good. Thus they can be used to get first reliable predictions
for possible cases not yet considered such as, for instance, the HFS of the $3s$ and $4s$ levels. 
They also give estimates of the $\texttt{D}_\texttt{21}$ values, with less accuracy but still with some meaningful
figures. These quantities, indeed, are differences of very nearby numbers and cannot be given any prevision using
the non relativistic Fermi energy.

We then consider in more detail the elementary process of a photon emission from a hyperfine transition of
the relativistic two fermion system. From the coupling of each component particle with an external
electromagnetic field, we determine the vertex in the global/relative coordinates and we calculate the transition rates.
We need, for this, the wave functions of the hyperfine states perturbatively corrected in the spin-spin interaction.
They are found  by evaluating the first order term of the Taylor series of the eigenspinors with respect to an external
parameter $\varepsilon$, namely by the same Hellmann-Feynman procedure used for the spectrum. In such a way, within the
numerical accuracy, we take into account the complete sum over all the intermediate states of the usual expansion
\cite{GS2}. We then carry on the elementary field theoretical calculation without any further
approximation avoiding, in particular, the evaluation of the (usually non relativistic) eigenfunctions at zero. We
recall that in the case of the quarkonium physics, this last procedure has been the source of an animated debate whether
a smoothing procedure at the origin was necessary or not \cite{RR, GS3}. Apart from this last minor point, however, the
correct covariant treatment of the two body fermion system naturally points out some features with a clear physical
explanation. In the first place the frequency of the emitted photon, determined by the four momentum conservation, shows
the recoil correction due to the finite mass of both components. Next, the matrix element of the process is given by the
sum of two contributions that involve the kinematical properties of each particle separately whose ratio turns out to be
the square of the ratio of the two component masses. Finally a new factor slightly modifying
the transition rate is generated by covariance. Due to the values of the masses involved in the systems under
investigation, all the previous corrections are indeed very small. 

In the last section we briefly summarize our results and we add a possible use of the explicit eigenfunctions to
calculate the one and two loops approximations of the electron vacuum polarization for the muonic Hydrogen, expressed by
means of the Uheling and  K\"allen-Sabry potentials \cite{UKSK,Pachu}. It is well known that they constitute the
dominant contributions to the Lamb shift for this atomic species \cite{Karshen}. We do not repeat, in this paper, the
derivation of the two fermion equation (\ref{BreitHam}), already exposed in \cite{GS1,GS2}. However, since the
framework is more general and changes have been brought with respect to what was presented in \cite{GS1,GS2}, we produce
all the necessary ingredients that make it possible to determine the actual system of equations and to calculate its
solutions. In order to maintain a plain exposition of the results they are gathered in the appendixes. In these we thus
explain the relationship between the two particle coordinates and the global/relative ones. We give the explicit
``16-dim spherical spinors'' obtained by diagonalizing angular momentum and parity of the two fermion system. We then
show the resulting eight independent first order equations and their reduction to the fourth order system
(\ref{system},\ref{even_coeff}) due to the existence of four algebraic relations among the unknown functions. This
procedure is completely analogous to the reduction of the Dirac equation in a central field. To conclude, we notice that
the charted course is absolutely straightforward and physically transparent, although its implementation has required a
certain deal of computer algebra, which also turned out to be rather useful for simplifying some thorny problems
connected with numerical precision.
%
%
\sect{The spectral problem for the HFS} \label{The_spectral_problem}
%
%
We denote by $\gamma_{{(i)}}$ the gamma
matrices acting in the spinor space of the $i$-th fermion of mass $m_{(i)}$. Assuming
 $m_{(1)}\geq m_{(2)}$ we put $M=m_{(1)}+m_{(2)}$ and $\rho=(m_{(1)}-m_{(2)})/M$.
In a system of units with $\hbar = c = 1$ and referring to Appendix A  for the coordinates in use,  
the two fermion wave equation reads \cite{GS1,GS2}
\begin{eqnarray}
&{}&\!\!\!\!\!\!\!\!\Bigl[\,\Bigl(
{\gamma}^0_{(1)}{\gamma_{(1)}}_{a}-
{\gamma}^0_{(2)}{\gamma_{(2)}}_{a}\Bigr)q_a+\frac12\Bigl({\gamma}^0_{(1)}\!
+\! {\gamma}^0_{(2)}\Bigr)M+\spazio{1.2}\cr
&{}&\!\!\!\frac12\Bigl({\gamma}^0_{(1)}\!
-\! {\gamma}^0_{(2)}\Bigr)M\rho -\Bigl(\lambda+\frac \alpha r\Bigr)\!+\! V_B(r)
\,\Bigr]\,\Psi(\emph{\textbf{r}})=0\,.
\label{BreitHam}
\end{eqnarray}
The eigenvalue $\lambda$ is the square root of the squared total momentum, $\alpha=e^2/\hbar c\equiv e^2$ is the
fine coupling constant, $\alpha/r$ the Coulomb interaction. Finally
\begin{eqnarray}
V_B(r)=\kappa_1\,\kappa_2\,\,\frac \alpha{2r}\,
{\gamma}^0_{(1)}{\gamma_{(1)}}_{a}{\gamma}^0_{(2)}{\gamma_{(2)}}_{b}
\Bigl(\delta_{ab}\!+\!\frac{r_ar_b}{r^2}\Bigr) 
\nonumber
\label{BreitTerm}
\end{eqnarray}
is the Breit term which describes the spin-spin interactions and generates the hyperfine splitting, $\kappa_1$ and
$\kappa_2$
being the factors accounting for the anomalous magnetic moments of the two fermions. As in \cite{GS2}, the first order
perturbative correction to the eigenvalues is evaluated by multiplying $V_B(r)$ times a parameter $\varepsilon$
in (\ref{BreitHam}) and taking the first term of the Taylor expansion of the eigenvalues with respect to $\varepsilon$
from the numerical solutions of the differential equations.
The radial system was obtained in \cite{GS1,GS2}. Introducing the dimensionless variables $s,\,w$ by 
\begin{eqnarray}
s = m_2\,r,\quad \lambda = m_2\,\Bigl(\,\frac{1+\rho}{1-\rho}+1+\frac{1+\rho}2\,\alpha^2 w\,\Bigl)
\label{dimensionless1}
\end{eqnarray}
we obtain for $\,m_2^{-1}(\lambda+\alpha/r)$ the dimensionless expression
\begin{eqnarray}
h(s) = \frac2{1-\rho}+\frac{1+\rho}2\,\alpha^2w+\frac\alpha s	\,. 
\label{dimensionless2}
\end{eqnarray}
The deduction of the system is briefly summarized in Appendix C in terms of
the spherical 16-components spinors, reported for completeness in in Appendix B. The general form
of the system  is therefore
\begin{eqnarray}
\frac{d}{ds}u_i(s) = \sum\limits_{j=1}^4 M_{ij}(s)\,u_j(s)
\label{system}
\end{eqnarray}
and the non vanishing elements of the matrix $M_{ij}(s)$ for the system with even parity are
\begin{eqnarray}
&{}& M_{{12}} \left( s \right)=M_{{43}} \left( s \right) ={\frac {2\rho\, \sqrt{j \left( j+1 \right) } }{
\left( 1-\rho \right) s\,h \left( s \right) }}\cr
&{}& M_{{13}} \left( s \right) =-\frac1{2h(s)}\, \Bigl(  h^2(s)-{\frac {4{\rho}^{2}}{
\left( 1-\rho \right) ^{2}}} \Bigr)\cr
&{}& M_{{21}} \left( s \right) =M_{{34}} \left( s \right) ={\frac { 2\rho\,\sqrt{j \left( j+1 \right) }}{ \left( 1-\rho
\right)  \left( s\,h
\left( s \right) -2\,\varepsilon\,\alpha \right) }}\cr
&{}& M_{{22}} \left( s \right) =M_{{44}} \left( s \right) =1/s \cr
&{}& M_{{24}} \left( s \right) = \frac{    s^2\,h^2(s)-{\displaystyle{\frac
{4{\rho}^{2} s^2}{ \left( 1-\rho \right) ^{2}}}} -4\,{\alpha}^{2}{\varepsilon}^{2}} { 2s\,(s\,h
\left( s \right) -2\,\varepsilon\,\alpha) }\cr
&{}& M_{{31}} \left( s \right) =\frac{h(s)}2 + \frac1{2s}\,\Bigl(
 4\,\varepsilon\,\alpha-{\frac {4j \left(
j+1 \right) }{s\,h \left( s \right) -2\,\varepsilon\,\alpha}}\cr
&{}&\phantom{ M_{{31}} \left( sxx \right)} -{\frac {4{s}^{2}}{ 
\left( 1-\rho \right) ^{2}(s\,h \left( s \right) - 4\,\varepsilon\,\alpha\,)   }}\,\Bigr)\cr
&{}& M_{{33}} \left( s \right) = 2/s\cr
&{}& M_{{42}} \left( s \right) =-M_{{24}} +{\frac {2j \left( j+1 \right)}{{s}^{2}\,h \left( s \right) }}
\label{even_coeff}
\end{eqnarray}
It is explained in Appendix C how to obtain the coefficients of the odd system from (\ref{even_coeff}).

We give some details concerning the numerical method we have used for the solution. We are dealing with a boundary value
problem for a fourth order differential system having two singular points, at the origin and at infinity.
No further singularities arise from the  matrix of the coefficients. The eigenvalues are found by means of a double
shooting method that produces a spectral equation from the vanishing of the determinant of the $4\!\times\!4$  of the
matching conditions at a chosen crossing point \cite{GS1, GS2, GS3}. The accuracy of the integrations can easily be kept
as high as needed. Approximate solutions at the origin and at infinity are necessary in order to start the numerical
integrations. This is perhaps the most delicate point of the whole procedure since the asymptotic solutions require a
high precision, checked and improved  by Pad\'e techniques. The numerical errors have been tested and estimated from the
stability of the results (spectral values, HFS, $\texttt{D}_\texttt{21}$ factors
and the other quantities presented in the following sections) \emph{vs.} the variations of the computational parameters
(initial points for the left and right integrations, accuracy of the asymptotic solutions, integration precision, choice
of the crossing point where the spectral equation is solved). We therefore consider meaningful the figures of the
results given in the Tables we present. For the proton, the electron and the muon we take from CODATA 2010 the
respective values: $\kappa_p=g_p/2=2.7928473565$, $\kappa_e=g_e/2=1.0011596522$, $\kappa_\mu=g_\mu/2= 1.0011659207$;
for the ${}^\texttt{3}\texttt{He}^+$ ion we have assumed the shielded Helion value $-3.1839627379413$ obtained from
$\kappa_p$, the ratio of the Helion to the proton mass equal to $2.9931526707$, the ratio of the shielded Helion to
proton magnetic moment ratio equal to $-0.761766558$ and taking into account the Helion atomic number.
Some further comments are in order. We stress again that all the nuclei of
our atoms have been assumed to be point-like and only their anomalous magnetic moments have been added. No other
nuclear properties have been considered even for the case of the ${}^3\texttt{He}^+$ ion. All the HFS have been
uniformly calculated by solving the spectral problem of equation (\ref{BreitHam}) in the $s$- and $p$-waves. This is
not the case when starting from a non relativistic approximation because of the vanishing of the $p$ wave functions at
the origin. The same procedure was used in \cite{GS3} to calculate the meson masses even in $d$- and $f$-wave, with an
excellent agreement with measurements, especially for heavy mesons. Here also we reproduce the available experimental
data  and most of the theoretical results
\cite{Karshenboim,Jungmann,Pachu,Martynenko} with an accuracy up to some units times $10^{-5}$. In particular the HFS
for the $1s$ and $2s$ levels of the muonic Hydrogen coincide with those commonly accepted within the
declared error. The same occurs for the $p^{1/2}$ and $p^{3/2}$ HFS of the muonium \cite{Jungmann}. A lower agreement,
instead, is observed for $\texttt{D}_{21}$ and for the $p$-wave HFS of the muonic Hydrogen. The latter difference
is mainly due to the initial contribution of the fourth order in the fine structure constant \cite{Martynenko}. An
analogous situation is met for the ${}^3\mathrm{He}^+$ ion \cite{Martynenko}, in contrast with the electronic Helium
ion, for which we reproduce quite well the HFS given in literature \cite{Karshen,Karshenboim} and  the
$\texttt{D}_{21}$ factor with an acceptable agreement. We will try to get a deeper insight into this problem in the
future.
\begin{table}[t]
{{ $~~~$
\begin{tabular}{lrrr}
  \hline
  {\texttt{Atom}}$\phantom{{}^{{}^{\displaystyle{i}}}}$ &$\Delta_\texttt{HFS}(\texttt{1s})$ 
  &$\Delta_\texttt{HFS}(\texttt{2s})$ &$ \texttt{D}_\texttt{21}~~~~$  
  \\
  \hline
  ({\texttt{p}},$\,${\texttt{e}})
&\texttt{1420.595}  
&\texttt{177.580}
&\texttt{0.04728}
 \\ 
  ({$\mu^+,$}$\,${\texttt{e}})
&\texttt{4464.481}  
&\texttt{558.078}
&\texttt{0.14617}
 \\ 
  ({${}^\texttt{3}\texttt{He}^+,$}$\,${\texttt{e}})
&\texttt{ -8665.637}  
&\texttt{ -1083.347}
&\texttt{ -1.14103}
 \\ 
 \hline
  ({\texttt{p}},$\,${\texttt{$\mu$}})
&\texttt{182.621}  
&\texttt{22.828}
&\texttt{0.00458}
 \\ 
  (${}^\texttt{3}\texttt{He}^+,$ $\,${\texttt{$\mu$}})
&\texttt{-1372.194}  
&\texttt{-171.544}
&\texttt{-0.16494}
 \\
 \hline
\end{tabular}
}}
\label{Table_D21}
\caption{The \texttt{1s}, \texttt{2s} HFS and the corresponding $ \texttt{D}_\texttt{21}$ for some Hydrogenic atoms
indicated, in the first column, by their
two components. $\texttt{p}$ is the proton, $\texttt{e}$ is the electron,
$\mu^+,\mu$ the positive and negative $\mu$ mesons, ${}^\texttt{3}\texttt{He}^+,$ the Helium 3 ion.
According to a common use, we give the results of the first three lines in MHz and those of the last two in
meV. Related data and results can be found in \cite{Karshenboim,Jungmann,Pachu,Martynenko}}.
\end{table}
%
%
%
%
\section{THE TRANSITION PROBABILITIES} \label{linewidth}
%
%
We retrace the general quantum mechanical procedure to calculate the transition probability between the
hyperfine split $s$-states. We consider in the Coulomb gauge the wave function of a photon with polarization
${\boldsymbol\epsilon}_\sigma$ \cite{Landau}:
$$ {\boldsymbol{A}}(k,\sigma)=
\frac{\sqrt{4\pi}}{\sqrt{2\omega}}\,\,{\boldsymbol\epsilon}_\sigma\,\,e^{\displaystyle -i\,kx}\,.$$
The interaction Hamiltonian reads
\begin{eqnarray}
H_{\mathrm{int}} = -e\Bigl(z\,\, \boldsymbol\alpha_{(1)}\!\cdot\!\emph{\textbf{A}}^{{(1)}}
-\boldsymbol\alpha_{(2)}\!\cdot\!\emph{\textbf{A}}^{{(2)}}\,\Bigr)
\nonumber
\end{eqnarray}
where $\emph{\textbf{A}}^{\textbf{(\textit{i})}}=\emph{\textbf{A}}(x_{(i)})$,
the charge of the lighter fermion (electron or $\mu$) is $-e$ and $z$ is the atomic number of the heavier fermion.
We choose the $\emph{\textbf{P}}_i=0$ frame and we use the coordinates $(Z,\emph{\textbf{r}})$ with the relations
(\ref{xi_di_Zr}) as explained in Appendix A. We factorize   the wave functions $\Psi_i(Z,\emph{\textbf{r}})$ and
$\Psi_f(Z,\emph{\textbf{r}})$  of initial and final atomic states, normalized in the box, into
\begin{eqnarray}
\Psi_{\ell}(Z,\emph{\textbf{r}}) = V^{-1/2}\,{e^{
-iP_{\ell}Z}}\,\psi_{\ell}(\emph{\textbf{r}})\,,\qquad  \ell=i,f\,.
\end{eqnarray}
where $\psi_i(\emph{\textbf{r}})$ and $\psi_f(\emph{\textbf{r}})$ are the 16-component spinors of Appendix B
corresponding to initial and final energies, angular momenta and parities. After some straightforward calculation and
using the definition of $\Delta$ given in Appendix A,  the first perturbation order of the $S$-matrix element  in the
finite volume $V$ normalization reads
\begin{eqnarray}
&{}& S_{fi} = -\frac{(2\pi)^4 ie}{\sqrt{2\omega V}}\,\,\delta^4(P_f+k-P_i) \int
d^3\emph{\textbf{r}}\,\,\frac {\sqrt{4\pi}}{\sqrt{V^2}}\,\psi^*_{f}(\emph{\textbf{r}})\,
\cr
&{}& \phantom{S_{fi}} \boldsymbol\epsilon_\sigma^*\!\cdot\!\Bigl[\,
\widetilde{\boldsymbol\alpha}_{(1)}
\,e^{-i(\frac12-\Delta){\emph{\textbf{k}}}\,\cdot\!\!\!\!{
\emph { \textbf { r }}}}
-\widetilde{\boldsymbol\alpha}_{(2)}\,\,e^{i(\frac12+\Delta){\emph{\textbf{k}}}\,
\cdot\!\!\!\! { \emph{ \textbf { r }}}}
\Bigr]\, \psi_{i}(\emph{\textbf{r}})\,.
\nonumber
\label{Smatrix}
\end{eqnarray}
Here $\widetilde{\boldsymbol\alpha}_{(j)}$ are the matrices obtained from ${\boldsymbol\alpha}_{(j)}$ by applying the
similarity transformation generated by the change of basis we have made so to give the spinor components the order
shown in Appendix B.
\begin{table}[t]
{{ $~~~$
\begin{tabular}{lrrrr}
  \hline
  {\texttt{Atom}}$\phantom{{}^{{}^{\displaystyle{i}}}}$ &$\Delta_\texttt{HFS}(\texttt{3s})$ 
  &$\Delta_\texttt{HFS}(\texttt{4s})$
&$\Delta_\texttt{HFS}(\texttt{2p}^\texttt{1/2})$  &$\Delta_\texttt{HFS}(\texttt{2p}^\texttt{3/2})$  
  \\
  \hline
  ({\texttt{p}},$\,${\texttt{e}})
&\texttt{52.617}  
&\texttt{22.198}
&\texttt{59.196}
&\texttt{23.678}
 \\ 
  ({$\mu^+,$}$\,${\texttt{e}})
&\texttt{165.357}  
&\texttt{69.760}
&\texttt{186.252}
&\texttt{74.629}
 \\ 
  ({${}^\texttt{3}\texttt{He}^+,$}$\,${\texttt{e}})
&$~$\texttt{-320.993}  
&$~$\texttt{-135.418}
&$~$\texttt{-361.100}
&\texttt{-144.385}
 \\ 
 \hline
  ({\texttt{p}},$\,${\texttt{$\mu$}})
&\texttt{6.764}  
&\texttt{2.854}
&\texttt{7.682}
&\texttt{3.115}
 \\ 
  ({${}^\texttt{3}\texttt{He}^+,$}$\,${\texttt{$\mu$}})
&\texttt{-50.828}  
&\texttt{-21.443}
&\texttt{-57.028}
&\texttt{-22.700}
 \\
 \hline
\end{tabular}
}}
\label{Table_34s2p}
\caption{
The \texttt{3s}, \texttt{4s}, $\texttt{2p}^\texttt{1/2}$, $\texttt{2p}^\texttt{3/2}$ HFS  for some
Hydrogenic atoms. Units are as in Table I. For data and results see \cite{Jungmann,Martynenko}}.
\end{table}
The $\delta^4$-function gives the energy-momentum conservation, 
$$P^0_i = P^0_f + \omega,\quad  \emph{\textbf{P}}_i = \emph{\textbf{P}}_f + \emph{\textbf{k}}$$ and contains the recoil
of the atom due to the radiation emission.When the radiation wavelength $2\pi/\omega$ is much larger than
the characteristic scale length of the atomic system, we can consider the first order expansion
$$e^{\pm i\,(\frac12\pm \Delta)\emph{\textbf{k}}\cdot\emph{\textbf{r}}} = 1\pm
\,(1/2\pm\Delta)\,\emph{\textbf{k}}\cdot\!\emph{\textbf{r}}+ o((\emph{\textbf{k}}\cdot\!\emph{\textbf{r}})^2)\,.
$$
The angular properties of the spinors imply that
$$\int d^3\emph{\textbf{r}}\,\,\,\psi^*_{f}(\emph{\textbf{r}})\, \widetilde{\boldsymbol\alpha}_{(j)}\,
\psi_{i}(\emph{\textbf{r}})
=0,\quad j=1,2. $$
We then let $\emph{\textbf{k}}=\omega\,\emph{\textbf{n}}\,$ with $\emph{\textbf{n}}^2=1$ and, taking into account the
$\delta^4$-function and the expression of $\Delta$ given in Appendix A, we introduce the quantity
\begin{eqnarray}
&{}&  \emph{\textbf{d}}_{fi} =-i\,e\int
d^3\emph{\textbf{r}}\,\,(\emph{\textbf{n}}\cdot\!\emph{\textbf{r}})\,\,\psi^*_{f}(\emph{\textbf{r}})\,\Bigl[\,\widetilde
{\boldsymbol\alpha}_{(1)}\,\Bigl(\frac12-\frac{m_1^2-m_2^2}{2\lambda_i^2}\Bigr) +\cr 
&{}&\phantom{\emph{\textbf{d}}_{fi} = -}
\widetilde{\boldsymbol\alpha}_{(2)}\,\Bigl(\frac12+\frac{m_1^2-m_2^2}{2\lambda_i^2}\Bigr)\,
\Bigr]\,\psi_{i}(\emph{\textbf{r}})\,.
\label{dipole}
\end{eqnarray}
The $S$-matrix element reads then
\begin{eqnarray}
 S_{fi} =
(4\pi\omega/
V^3)^{1/2}\,(2\pi)^4\,\delta^4(P_f+k-P_i)\,\,(\boldsymbol\epsilon_\sigma^*\!\cdot\!\emph{\textbf{d}}_{fi})
\nonumber
\end{eqnarray}
and, as usual,  we get a transition rate 
\begin{eqnarray}
dw =
\frac\omega{2\pi}\,\delta^4(P_f+k-P_i)\sum\limits_{\sigma}|\boldsymbol\epsilon_\sigma^*\!\cdot\!\!\emph{\textbf{
d}}_{fi}|^2\,\,{d^3\!\!\emph { \textbf{k}}}\,\,{d^3\emph{\textbf{P}}_f}
\nonumber
\end{eqnarray}
Recalling that
$\int{d^3\emph{\textbf{P}}}/{2P^0} = \int d^4{P}\,\theta(P^0)\,\delta(P^2-\lambda^2)$
we next integrate over the final global momentum, finding
\begin{eqnarray}
\frac{dw}{d\omega\,d\Omega_n}\! = \!\frac{\omega^3
}{2\pi\lambda_i}\,
(\lambda_i-\omega)\,\delta\Bigl(\omega-\frac{\lambda_f^2-\lambda_i^2}{2\lambda_i}\Bigr)\sum\limits_{\sigma}
|\boldsymbol\epsilon_\sigma^*\!\cdot\!\emph{ \textbf {d}}_{fi}|^2 
\nonumber
\end{eqnarray}
where $d\Omega_n$ is the unit solid angle in the direction $\emph{\textbf{n}}$.
Reinserting the $\hbar$ and $c$ factors, the final integration over the solid angle gives the total transition rate
\begin{eqnarray}
w = \frac 43\,\frac{\omega^3}{\hbar c^3}
\,\Lambda_{fi}^2\,|\, {{\mu}_{fi}}\,|^2 
\label{w}
\end{eqnarray}
where
$$\hbar\omega= \frac{\lambda_i+\lambda_f}{2\lambda_i}\,(\lambda_i-\lambda_f)\,,\qquad\Lambda_{fi}^2 =
\frac{\lambda^2_i+\lambda^2_f}{2\lambda^2_i}\,$$
and $|\, {{\mu}_{fi}}\,|^2$ is the common value of $|\boldsymbol\epsilon_\sigma^*\!\cdot\!\emph{ \textbf
{d}}_{fi}|^2$ for each of the two independent circular polarizations
$(\boldsymbol{\epsilon}_1\pm i\boldsymbol{\epsilon}_2)/\sqrt{2}$.

Observe that the two different contributions present in the expression (\ref{dipole}) -- and consequently in
the transition rate $w$ -- clearly reflect the fact that we are dealing with a genuine two body problem. Indeed,
loosely speaking, the two terms can be attributed to the two different fermions, whose velocities are
represented by the $\boldsymbol\alpha$ matrices of the corresponding tensor component. The numerical results confirm
this interpretation showing that the ratio of the two contribution turns out to be practically coincident with the ratio
of the square of the fermion masses.

Let us now consider in more detail the $\emph{\textbf{d}}_{fi}$ for the $s$ hyperfine transitions. We have
taken, for computational convenience, the spinors $\psi_{i}(\emph{\textbf{r}})$ referring to the $n1s0$ states and the
spinors $\psi_{f}(\emph{\textbf{r}})$ to the $n3s1$ states, $n$ denoting the level quantum number. For
the $n3s1$ states, having $j=1$, the value of $m=-1,0,1$ has
also to be specified. Recalling that 
\begin{eqnarray}
&{}&
 \emph{\textbf{n}}\cdot\!\emph{\textbf{s}}=\sqrt{\frac{2\pi}{3}}\,s\,\Bigl(\,Y^1_{-1}(\theta,\phi)\,n_+ -
Y^1_{1}(\theta,\phi)\,n_-\cr
&{}&\phantom{\emph{\textbf{n}}\cdot\!\emph{\textbf{s}}=XXXXX s\Bigl(Y^1_{-1}(\theta,\phi)\,n_+}
+\sqrt{2}\,
Y^1_{0}(\theta,\phi)\,n_3\,\Bigr),
\nonumber
\end{eqnarray}
in the dimensionless variables (\ref{dimensionless1}), 
the matrix element (\ref{dipole}) is given by combinations of the integrals
\begin{eqnarray} 
D^{(j)}_{[n,m,a,b]}=\int\,d^3\!\emph{\textbf{s}}
\,\,\psi_{n3s1,\,m}^*(s)\,\,Y^1_{a}(\theta,\phi)\,s\,\,\widetilde{\alpha}_{(j),b}\,\,\psi_{n1s0}(s)
\nonumber
\end{eqnarray}
with $m,a=-1,0,1$;   $j=1,2$; $b=1,2,3$.
For a given $m$ only few components of the
$\widetilde{\alpha}$-matrices give non vanishing contribution. Moreover, integrating over the angular variables, the
spherical symmetry gives rise to identities reducing the problem to the calculation of a single integral for each
tensor type of $\widetilde{\alpha}$-matrix. We have indeed
\begin{eqnarray} 
&{}& D^{(j)}_{[n,1,0,1]\,} = \phantom{-}iD^{(j)}_{[n,1,0,2]} = D^{(j)}_{[n,-1,0,1]\phantom{-}} =
-iD^{(j)}_{[n,-1,0,2]}=\cr
&{}& D^{(j)}_{[n,0,1,1]\,} = -iD^{(j)}_{[n,0,1,2]} = D^{(j)}_{[n,0,-1,-1]}=
\phantom{-}iD^{(j)}_{[n,0,-1,2]}=\cr
&{}& \frac{D^{(j)}_{[n,1,1,3]}}{\sqrt{2}}= -\frac{D^{(j)}_{[n,-1,-1,3]}}{\sqrt{2}}
\nonumber
\end{eqnarray}
all other choices of $m,a,b$ giving a vanishing result.

After the integration over the angular variables, we are finally left with
\begin{eqnarray}
&{}& D^{(1)}_{[n,1,0,1]} = -\frac1{12\sqrt{\pi}}\int\limits_0^\infty ds\,s^3 \Bigl[
\sqrt{6}\Bigl(d_0^{(s0)}c_1^{(s1)}+d_1^{(s0)}c_0^{(s1)}\Bigr)+\spazio{1.0}\cr
&{}&\sqrt{3}\Bigl(d_0^{(s0)}d_1^{(s1)}+d_1^{(s0)}d_0^{(s1)}\Bigr)
 +3\Bigl(a_0^{(s0)}b_1^{(s1)}+a_1^{(s0)}b_0^{(s1)}\Bigr)\,\Bigr],\spazio{1.7}\cr
&{}& D^{(2)}_{[n,1,0,1]} = -\frac1{12\sqrt{\pi}}\int\limits_0^\infty ds\,s^3 \Bigl[
\sqrt{6}\Bigl(d_0^{(s0)}c_0^{(s1)}+d_1^{(s0)}c_1^{(s1)}\Bigr)+\spazio{1.0}\cr
&{}&
\sqrt{3}\Bigl(d_0^{(s0)}d_0^{(s1)}+d_1^{(s0)}d_1^{(s1)}\Bigr) -3\Bigl(a_0^{(s0)}b_0^{(s1)}
+a_1^{(s0)}b_1^{(s1)}\Bigr)\,\Bigr].
\nonumber
\end{eqnarray}
where we have denoted by $a_i^{(s0)},d_i^{(s0)}$, $i=0,1$, the coefficients  $a_i{(s)},\,d_i{(s)}$ of the
spinors of the states $n1s0$ normalized to unity and by  $b_i^{(s1)},c_i^{(s1)},d_i^{(s1)}$,
$i=0,1$, the coefficients $b_i{(s)},\,c_i{(s)},\,d_i{(s)}$ of the spinors of the states $n3s1$ with
$m=1$ normalized to unity (see Appendix B).
\begin{table}[t]
{{ $~~~$
\begin{tabular}{lcccc}
  \hline
  {\texttt{Atom}}$\phantom{{}^{{}^{\displaystyle{i}}}}$
&$\texttt{10}^{\texttt{4}}\,\texttt{D}^{(\texttt{1})}_{[\texttt{1101}]}$ 
 &$\texttt{D}^{(\texttt{2})}_{[\texttt{1101}]}$
&$\texttt{10}^{\texttt{4}}\,\texttt{D}^{(\texttt{1})}_{[\texttt{2101}]}$
&$\texttt{D}^{(\texttt{2})}_{[\texttt{2101}]}$ 
  \\
  \hline
  ({\texttt{p}},$\,${\texttt{e}})
&\texttt{.9408482 }  
&\texttt{.1727440 }
&\texttt{.9408195 }
&\texttt{.1727463 }
 \\ 
  ({\texttt{p}},$\,${\texttt{$\mu$}})
&\texttt{194.5346 }  
&\texttt{.1727434 }
&\texttt{194.5309 }
&\texttt{.1727462 }
 \\ 
 \hline
\end{tabular}
}}
\label{Table_Dmab12}
\caption{Numerical results for some $\texttt{D}^\texttt{(j)}_\texttt{[n,m,a,b]}\,$.}
\end{table}
Since the hyperfine splitting is due to the Breit term, the eigenfunctions of the split hyperfine states make sense only
if they include the first perturbative correction due to the Breit interaction. It is equally evident that  in practice
the correction cannot be calculated by an expansion over a complete set of states. As we said in the introduction,
instead, we adopt for the eigenfunctions the same procedure used for the spectrum: namely, we calculate the
solutions of the systems (\ref{system}) for $\varepsilon V_B(r)$,  for different values of $\varepsilon$, at the
corresponding eigenvalues,  we construct the spinors of Appendix B and we find the corrected eigenstates by means of a
first order expansion in $\varepsilon$. A detailed proof of the procedure is given in \cite{GS2}. In Table III we report
the numerical results for the ${D}^{({j})}_{[{n101}]}$, $n,j=1,2$,  of the electronic and muonic Hydrogen. Finally,
after the insertion of the appropriate $\hbar$ and $c$ factors, we find that $\mu_{fi}= 9.273845
~\mathrm{erg}\,\mathrm{Gauss}^{-1}\,,$ a value very close to the Bohr magneton commonly used for this type of
calculation. The resulting transition rate turns out to be $\simeq\,\,2.87 \times 10^{-15}~ \mathrm{sec}^{-1}$.

Let us focus on some features cleanly pointed out by the completely covariant treatment of the photon emission  by a
two fermion system that we have presented. In the first place we look at the emitted radiation and
we find separate contributions due to the two particles, corresponding to the terms
$\widetilde{\boldsymbol\alpha}_{(1,2)}\,[1/2\pm (m_1^2-m_2^2)/(2\lambda_i^2)]$. From Table III we see that
${D}^{({2})}_{[{1101}]}/{D}^{({1})}_{[{1101}]}=1836.04$ and ${D}^{({2})}_{[{1101}]}/{D}^{({1})}_{[{1101}]}=8.88$ for the
$n=1$ hyperfine transition of the electronic and muonic Hydrogen respectively, reproducing very accurately the ratios of
the proton mass to the electron and muon masses. On the other hand $\Delta \simeq (m_1^2-m_2^2)/2(m_1+m_2)^2 \simeq
1/2-m_2/m_1$, so that $1/2+\Delta\simeq1$ and $1/2-\Delta\simeq m_2/m_1$, giving a ratio of the two amplitudes
very closely proportional to $(m_1/m_2)^2$.
Next we observe that the emission process determines the photon frequency carrying a correction
$(\lambda_i+\lambda_f)/2\lambda_i$ to the pure spectroscopic difference of the energy of
the levels. Its origin is evident from the dynamics of the process. In the cases we have dealt with, this correction is
very tiny: even in the muonic Hydrogen, where the effect is larger, the only appreciable change is for the level $21s0$
which passes from 182.621 to 182.614 meV. Finally, a further factor $\Lambda_{fi}^2$ correcting almost imperceptibly the
transition rate arises from the covariance.


\section{CONCLUSIONS} \label{conclusions}


\begin{table}[b]
{{ $~~~$
\begin{tabular}{lrrr}
  \hline
  {\texttt{}}$\phantom{{}^{{}^{\displaystyle{i}}}}$
&$     \Delta \texttt{E}\texttt{(}\texttt{V}_\texttt{1}\texttt{)}$ 
&$     \Delta \texttt{E}\texttt{(}\texttt{V}_\texttt{11}\texttt{)}$ 
&$     \Delta \texttt{E}\texttt{(}\texttt{V}_\texttt{2}\texttt{)}$ 
  \\
  \hline
  \texttt{2s}
&\texttt{  219.589}  
&\texttt{    0.248881}
&\texttt{     1.343129}
 \\ 
  \texttt{2p}
&\texttt{  14.581}  
&\texttt{  -0.000776}
&\texttt{    0.051235}
 \\ 
$\Delta_{\texttt{2s}-\texttt{2p}}$
&\texttt{  205.008}  
&\texttt{   0.249657}
&\texttt{   1.291894}
 \\ 
 \hline
\end{tabular}
}}
\label{Lamb}
\caption{Energy shifts due to the electron vacuum polarization in meV.
We have omitted the much smaller contribution of the 1P-reducible two loop diagram.}
\end{table}
In this final section we  summarize the results we have found so far and we add some comments on the points we think to
be the most relevant. Firstly we stress again that the covariant treatment of two relativistic interacting fermions
necessarily implies a fourth order system. The attempt of using second order differential equations that occurs, for
instance, for a non relativistic initial description of the atom invariably results in a perturbation series
not very easy to deal concurrently with the series of the radiative corrections. On the one hand, the analytic
discussion of the  equations becomes exceedingly hard and we have to resort to a numerical treatment if we want a non
perturbative solution. The method applies uniformly to all the atomic
states so that we can reasonably expect the same order of accuracy for the spectral levels  provided that the numerical
precision is maintained sufficiently high.  On the other hand, our approach also permits a
straightforward determination of the wave functions. When it is required, the first order perturbation correction due to
the spin-spin interaction is calculated easily enough by a numerical treatment. The
eigenfunctions have been used to find in a completely covariant way the emission rate of a
photon in a hyperfine transition. To conclude, we want to present the results obtained by using our relativistic purely
Coulomb eigenfunctions to determine the corrections to the $2s$ and $2p$ levels of the muonic Hydrogen due to the vacuum
polarization by electrons \cite{UKSK,Pachu,Karshen} that constitute the leading
contribution to the Lamb shift for the muonic Hydrogen. We refer to \cite{Karshen} for the Feynman diagrams they are
related to. The shifts to the energy we have found  are given in Table IV and they are in complete agreement with those
reported in literature. We finally notice that in
treating the hyperfine transition we have, in a certain sense, determined the interaction vertex of a photon with the
composite two fermion system. We would naturally carry on with the electrodynamical radiative corrections in our
covariant framework. For this purpose the propagator of the two interacting fermions by the
Coulomb potential is required. Obviously its determination cannot be other than numerical and work is in progress in
this direction.


\section*{APPENDIX A. COORDINATES} \label{appendixa}


Denote by  $x^\mu_{(i)}$ and $p^\mu_{(i)}$
coordinates and momenta of the two fermions. We indicate in boldface the spatial 3-vectors. 
Denote also by $\gamma^\mu_{(1)}=\gamma^\mu\otimes \mathbb{I}$ and  $\gamma^\mu_{(2)}=\mathbb{I}\otimes\gamma^\mu$
the $\gamma$-matrices acting on the spinor components of the first and second particle respectively. The same
notation is used for the $\alpha$-matrices.
Let $\eta_{\mu\nu}$ be the Minkowski metric and
\begin{eqnarray}
\label{Xpr}
&{}&\!\!\!\!\!\!\!\!X^\mu=(1/2) \,(x^\mu_{(1)}+x^\mu_{(2)})\,~~~P^\mu=p^\mu_{(1)}+p^\mu_{(2)}\cr
&{}&\!\!\!\!\!\!\!\!r^\mu\,=x^\mu_{(1)}-x^\mu_{(2)}\,~~~~~~~~~~\phantom{\frac12~}q^\mu=(1/2)
\,(p^\mu_{(1)}-p^\mu_{(2)})\, .
\nonumber
\end{eqnarray}
We define the tensor
\begin{eqnarray}
\label{varepsilon_mu_nu}
\varepsilon_0^\mu(P)=\,\frac{P^\mu}{\sqrt{P^2}}\,,\quad
\varepsilon_a^\mu(P)=\eta_a^\mu-\displaystyle{\frac{P_a\,[\,P^\mu
+\eta_0^\mu\sqrt{P^2}\,]}{\sqrt{P^2}\,[\,P_0+\sqrt{P^2}\,]}}\,
\nonumber
\end{eqnarray}
$(a=1,2,3)$. It represents a Lorentz transformation to the
$\emph{\textbf{P}}=0$ frame as it satisfies the identities
{{
\begin{eqnarray}
\!\!\!\!\! \eta_{\mu\nu}\,\varepsilon_\alpha^\mu(P)\,
\varepsilon_\beta^\nu(P)\,= \,\eta_{\alpha \beta}\,,~~~
\eta_{\alpha \beta}\,\varepsilon_\alpha^\mu(P)\,
\varepsilon_\beta^\nu(P)\,= \,\eta^{\mu\nu}\nonumber
\end{eqnarray}
}}
 We construct the canonical variables
{{
\begin{eqnarray}
&{}& Z^\mu\!=\!X^\mu\!+\!{\frac
{\varepsilon_{abc}P_a \eta_b^\mu L_c}{\sqrt{P^2}[P_0+
\sqrt{P^2}]}}
\!+\!{\frac{\varepsilon_a^\mu}{\sqrt{P^2}}}\ta
q_a\breve r-{r_a\breve q}\tc
\!+\!{\frac{P^\mu}{P^2}}\breve q\breve r
\spazio{1.4}\cr 
&{}& \breve{q}=\varepsilon^\mu_0\,
q_\mu\,,~~~\, \breve{r}=\varepsilon^\mu_0 r_\mu\,,~~~
q_a=\varepsilon_a^\mu\, q_\mu\,,~~~ r_a=\varepsilon_a^\mu r_\mu\,.
\nonumber
\label{variabili}
\end{eqnarray}
}}
where $\emph{\textbf{r}}$ and $\emph{\textbf{q}}$ are Wigner vectors of spin one and $\emph{\textbf{Z}}$
is a Newton-Wigner position vector for a particle with angular
momentum ${L_a}=\varepsilon_{abc}\,\,r_b\, q_c\,$. In \cite{GS1,GS2} a canonical reduction of the phase space
was obtained by $\breve{q}=\sqrt{P^2}\Delta$, with $\Delta=(m_{1}^2-m_{2}^2)/(2{P^2})$, corresponding to the
cyclic relative time coordinate $\breve{r}$. This is the phase space where interactions are defined. In the
$\emph{\textbf{P}}=0$ frame we then have 
\begin{eqnarray}
\emph{\textbf{x}}_{(1)}=\emph{\textbf{Z}}+(1/2-\Delta)\emph{\textbf{r}}\,,~~
\emph{\textbf{x}}_{(2)}=\emph{\textbf{Z}}-(1/2+\Delta)\emph{\textbf{r}}\,.~~~
\label{xi_di_Zr}
\end{eqnarray}
where now $\Delta=(m_{1}^2-m_{2}^2)/2\lambda^2$. 


\section*{APPENDIX B. SPHERICAL 16-DIM SPINORS} \label{appendixb}


\renewcommand{\theequation}{A.\arabic{equation}}
\setcounter{equation}{0}

For the sake of completeness we report here the  state vectors $\Psi_{+}$, $\Psi_{-}$
of definite energy, angular momentum $(j,m)$, even and odd
parity $(-)^j$ and $(-)^{j+1}$with respect to the angular momentum. The even state  is given the order 
\begin{eqnarray}
&{}&\Psi_{+}={}^t\Bigl(\,
\Psi_{+}^{(M)},\,
\Psi_{+}^{(-M)},\,
\Psi_{+}^{(-\mu)},\,
\Psi_{+}^{(\mu)}
\,\Bigr)\quad~~ \mathrm{where}\cr
&{}&\Psi_{+}^{(Q)}={}^t\Bigl(\,
\psi_{+\,0}^{(Q)}\,,\,
\psi_{+\,1_+}^{(Q)}\,,\,
\psi_{+\,1_0}^{(Q)}\,,\,
\psi_{+\,1_-}^{(Q)}
\,\Bigr)\,,~~Q=\pm M\,,\,\mp\mu\,.
\nonumber
\label{StatoPari}
\end{eqnarray}
The explicit expressions of the components read: 
\begin{eqnarray} 
%
%
%
&{}& \psi_{+\,0}^{(M)}= Y^{j}_{m}(\theta,\,\phi)\,{a_{0}}(s)
~~~~~~~~~\, 
\spazio {1.2}  \cr
%
%
&{}& \psi_{+\,1_+}^{(M)}= - 
{\displaystyle \frac {\sqrt{j - m + 1}\sqrt{j + m}}{\sqrt{2\,j}
\sqrt{j+1}}}\,
Y^{j}_{m-1}(\theta,\,\phi)\,{b_{0}}(s)   
\spazio {1.2}  \cr
%
%
&{}&\psi_{+\,1_0}^{(M)}= {\displaystyle \frac {m}{\sqrt{j}\,\sqrt{1 + j}}}
\,Y^{j}_{m}(\theta,\,\phi)\,
\,{b_{0}}(s) 
\spazio {1.2}  \cr
%
%
&{}&\psi_{+\,1_-}^{(M)}= 
{\displaystyle \frac {
\sqrt{j - m}\,\sqrt{j + m + 1}}{\sqrt{2\,j}
\sqrt{j+1}}}\,
Y^{j}_{m+1}(\theta,\,\phi)\,{b_{0}}(s) 
\spazio {1.2}  \nonumber\cr
%
%
%
%
&{}& \psi_{+\,0}^{(-M)}=Y^{j}_{m}(\theta,\,\phi)\,{a_{1}}(s)
\spazio {1.2}  \cr
%
%
&{}& \psi_{+\,1_+}^{(-M)}=- {\displaystyle 
\frac {\sqrt{j - m + 1}\sqrt{j + m}}{\sqrt{2\,j}\sqrt{
 j+1}}}\,
Y^{j}_{m-1}(\theta,\phi)\,{b_{1}}(s)    
\spazio {1.2}  \cr
%
%
&{}&\psi_{+\,1_0}^{(-M)}={\displaystyle \frac {m}{\sqrt{j}\,\sqrt{1 + j}}}
\,Y^{j}_{m}(\theta,\,\phi)
\,{b_{1}}(s)
\spazio {1.2}  \cr
%
%
&{}&\psi_{+\,1_-}^{(-M)}= {\displaystyle \frac {\sqrt{j - m}\sqrt{j + m + 1}}{\sqrt{2\,j}
\sqrt{j+1}}}\,
Y^{j}_{m+1}(\theta,\phi)\,{b_{1}}(s)   
\spazio {1.2}  \nonumber\cr
%
%
%
%
%
%
&{}& \psi_{+\,0}^{(-\mu)}=0
\spazio {1.2}  \cr
%
%
 &{}& \psi_{+\,1_+}^{(-\mu)}={\displaystyle 
\frac {\sqrt{j + m
 - 1}\,\sqrt{j + m}}{\sqrt{2\,j}\,\sqrt{2\,j - 1}}}\,
 Y^{j-1}_{m-1}(\theta,\,\phi)\,
 {c_0}(s)\spazio {1.0}\cr 
&{}&  \phantom{XX}+{\displaystyle \frac {\sqrt{j - m + 1}\,
 \sqrt{j - m + 2
}}{
\sqrt{ 2\,j+2}\,\sqrt{ 2\,j+3}}}
\,Y^{j+1}_{m-1}(\theta,\,\phi)\,{d_{0}}(s) 
\spazio {1.2}  \cr
%
%
&{}&\psi_{+\,1_0}^{(-\mu)}={\displaystyle \frac {\sqrt{j - m}\,\sqrt{j + m}}{\sqrt{j}\,
\sqrt{2\,j
 - 1}}}\,
Y^{j-1}_{m}(\theta,\,\phi)\,
{c_0}(s)\spazio {1.0}\cr 
&{}& \phantom{XX}- {\displaystyle \frac {\sqrt{j - m + 1}\,
 \sqrt{j + m + 1
}}{\sqrt{
1 + j}\,\sqrt{ 2\,j+3}}}\,Y^{j+1}_{m}(\theta,\,\phi)\,{d_{0}}(s)  
\spazio {1.2}  \cr
%
%
&{}&\psi_{+\,1_-}^{(-\mu)}={\displaystyle \frac {\sqrt{j - m
 - 1}\,\sqrt{j - m}}{\sqrt{2\,j}\,\sqrt{2\,j - 1}}}\,
 Y^{j-1}_{m+1}(\theta,\,\phi)\,
 {c_0}(s)\spazio {1.0}\cr
&{}& \phantom{XX}+ {\displaystyle \frac {\sqrt{j + m + 1}\,
 \sqrt{j + m + 2
}}{
\sqrt{ 2\,j+2}\,\sqrt{ 2\,j+3}}}
\,Y^{j+1}_{m+1}(\theta,\,\phi)\,{d_{0}}(s)   
\spazio {1.2}  \nonumber\cr
%
%
%
%
&{}& \psi_{+\,0}^{(\mu)}=0  
\spazio {1.2}  \cr
%
%
&{}&\psi_{+\,1_+}^{(\mu)}={\displaystyle 
\frac {\sqrt{j + m
 - 1}\,\sqrt{j + m}}{\sqrt{2\,j}\,\sqrt{2\,j - 1}}}\,
 Y^{j-1}_{m-1}(\theta,\,\phi)\,
 {c_{1}}(s)\spazio {1.0}\cr 
&{}&  \phantom{XX}+ {\displaystyle \frac {\sqrt{j - m + 1}\,
\sqrt{j- m + 2
}}{
\sqrt{ 2\,j+2}\,\sqrt{ 2\,j+3}}}
\,Y^{j+1}_{m-1}(\theta,\,\phi)\,{d_{1}}(s)   
\spazio {1.2}  \cr
%
%
 &{}&\psi_{+\,1_0}^{(\mu)}={\displaystyle \frac {\sqrt{j - m}\,\sqrt{j + m}}{\sqrt{j}
 \,\sqrt{2\,j - 1
}}}\,
 Y^{j-1}_{m}(\theta,\,\phi)\,
 {c_{1}}(s)\spazio {1.0}\cr 
&{}&  \phantom{XX}- {\displaystyle \frac {\sqrt{j - m + 1}\,
 \sqrt{j + m + 1
}}{\sqrt{
j + 1}\,\sqrt{ 2\,j+3}}}
\,Y^{j+1}_{m}(\theta,\,\phi)\,{d_{1}}(s)   
\spazio {1.2}  \cr
%
%
&{}&\psi_{+\,1_-}^{(\mu)}={\displaystyle 
\frac {\sqrt{j - m
 - 1}\,\sqrt{j - m}}{\sqrt{2\,j}\,\sqrt{2\,j - 1}}}\,
 Y^{j-1}_{m+1}(\theta,\,\phi)\,
 {c_{1}}(s)\spazio {1.0}\cr 
&{}&  \phantom{XX}+ {\displaystyle \frac {\sqrt{j + m + 1}\,
\sqrt{j + m + 2
}}{
\sqrt{ 2\,j+2}\,\sqrt{ 2\,j+3}}}
\,Y^{j+1}_{m+1}(\theta,\,\phi)\,{d_{1}}(s)
\nonumber
%
%
%
%
\label{psipari16}
\end{eqnarray}
We get the odd state $\Psi_{-}$ by the parity transformation
%
$$
\Psi_{-}=
%
%
%
%
\left(\,
\begin{matrix}
0 & {\mathbb I}_8  \spazio{0.6}\cr
 {\mathbb I}_8 & 0 
 \end{matrix}
 \,\right)\,\Psi_{+}
%
%
%
\label{StatoDispari}
$$
%
This amounts to changing the sign of the mass $m_{1}$.
%


\section*{APPENDIX C. EQUATIONS} \label{appendixc}


The radial differential equations are obtained by applying the Hamiltonian operator (\ref{BreitHam})
to the even and odd states and requiring the vanishing of the coefficients of the different
spherical harmonics in each component of the resulting vector. For each parity we get eight independent
equations.
Let $D_s^{[\kappa]} = d/ds + \kappa/s$ and  $f_\pm(s) = f_0(s)\pm f_1(s)\,$, $~f=a,b,c,d$.  
The even system is:
\begin{eqnarray}
%
%
&{}&\sqrt{j}\,D_s^{[j+1]}\,a_+(s) - \sqrt{j + 1}\,D_s^{[j+1]}\,b_-(s)\cr
&{}& \phantom{\sqrt{j}D_s}
+\sqrt{2j+1}\,c_0(s)\Bigl(\frac{2\rho}{1-\rho}+h(s)\Bigr)\cr
&{}& \phantom{\sqrt{j}D_s}-\frac{2\varepsilon\alpha(j+1)c_1(s)}{s\sqrt{2j+1}}-\frac{2\varepsilon\alpha
\sqrt{j(j+1)}\,d_1(s)}{s\sqrt{2j+1}}=0
 \spazio{1.}\cr
%
%
&{}&\sqrt{j}\,D_s^{[j+1]}\,a_+(s) + \sqrt{j + 1}\,D_s^{[j+1]}\,b_-(s)\cr
&{}& \phantom{\sqrt{j}D_s}
-\sqrt{2j+1}\,c_1(s)\Bigl(\frac{2\rho}{1-\rho}-h(s)\Bigr)\cr
&{}& \phantom{\sqrt{j}D_s}
-\frac{2\varepsilon\alpha(j+1)c_0(s)}{s\sqrt{2j+1}} -\frac{2\varepsilon\alpha\sqrt{j(j+1)}d_0(s)}{s\sqrt{2j+1}}=0
 \spazio{1.}\cr
%
%
&{}&\sqrt{j + 1}\,D_s^{[-j]}\,a_+(s) + \sqrt{j}\,
D_s^{[-j]}\,b_-(s)\cr
&{}& \phantom{\sqrt{j}D_s}-\sqrt{2 j+1} d_0(s) \Bigl(\frac{2 \rho}{1-\rho}+h(s)\Bigr)\cr
&{}& \phantom{\sqrt{j}D_s}+\frac{2 \varepsilon \alpha j
d_1(s)}{s \sqrt{2 j+1}}+\frac{2 \varepsilon \alpha 
\sqrt{j(j+1)} c_1(s)}{s \sqrt{2 j+1}}=0
\spazio{1.}\cr
%
%
&{}&\sqrt{j + 1}\,D_s^{[-j]}\,a_+(s)  - \sqrt{j}\,
D_s^{[-j]}\,b_-(s)\cr
&{}& \phantom{\sqrt{j}D_s} + \sqrt{2 j+1} d_1(s) \Bigl(\frac{2 \rho}{1-\rho}-h(s)\Bigr)\cr
&{}& \phantom{\sqrt{j}D_s}+\frac{2 \varepsilon \alpha j
d_0(s)}{s \sqrt{2 j+1}}+\frac{2 \varepsilon \alpha  \sqrt{j(j+1)} c_0(s)}{s  \sqrt{2 j+1}}=0
\spazio{1.}\cr
%
%
&{}&\sqrt{j}\,D_s^{[-j+1]}\,c_+(s)
- \sqrt{j + 1}\,D_s^{[j+2]} \,d_+(s)\cr
&{}&\,+\sqrt{2 j+1} \,a_0(s) \Bigl(\frac 2{1-\rho}-h(s)\Bigr)
-\frac{4 \varepsilon \alpha \sqrt{2j+1}\, a_1(s)}s=0
 \spazio{1.}\cr
%
%
&{}&\sqrt{j}\,D_s^{[-j+1]}\,c_+(s)
- \sqrt{j + 1}\,D_s^{[j+2]}\,d_+(s)\cr
&{}&\, - \sqrt{2 j+1}\, a_1(s) \Bigl(\frac2{1-\rho}+h(s)\Bigr)-\frac{4 \varepsilon \alpha  \sqrt{2 j+1}\, a_0(s)}s=0
 \spazio{1.0}\cr
%
%
&{}&\sqrt{j + 1}\,D_s^{[-j+1]} \,c_-(s)
+ \sqrt{j}\,D_s^{[j+2]}\,d_-(s)\cr
&{}&\, -\sqrt{2 j+1}\, b_0(s) \Bigl(\frac 2{1-\rho}-h(s)\Bigr)-\frac{2 \varepsilon \alpha \sqrt{2 j+1}\, b_1(s)}s=0
\spazio{1.0}\cr
%
%
&{}&\sqrt{j + 1}\,D_s^{[-j+1]} \,c_-(s)
+ \sqrt{j}\,D_s^{[j+2]}\,d_-(s)\cr
&{}&\, -\sqrt{2 j+1}\, b_1(s) \Bigl(\frac2{1- \rho}+h(s)\Bigr)+\frac{2 \varepsilon \alpha \sqrt{2 j+1} b_0(s)}s =0
\nonumber
\end{eqnarray}
From it we get four algebraic equation. Defining
\begin{eqnarray}
&{}&{\it u_\pm(s)}=-{\frac { \sqrt{j}{\it c_\pm(s)}- \sqrt{j+1}{\it d_\pm(s)}}{\sqrt{2\,j+1}}}\,\cr
&{}&{\it v_\pm(s)}=-{\frac {\sqrt{j+1}{\it c_\pm(s)}+ \sqrt{j}{\it d_\pm(s)}}{ \sqrt{2\,j+1}}}\,
\nonumber
\end{eqnarray}
and introducing $u_i(s),\, i=1,..,4$, with $u_1(s)=a_+(s)$, $u_2(s)=b_-(s)$, $u_3(s)=u_+(s)$, $u_4(s)=v_-(s)$, they
read
\begin{eqnarray}
&{}&{\it a_-}(s) ={\frac {2s\,u_{{1}} \left( s \right)}{ \left( 1-\rho \right)  \left( s\,h
\left( s \right) -4\,\alpha\,\varepsilon \right)}}
\spazio{0.8}\cr
&{}&
{\it b_+} \left( s \right) ={\frac {2s\,u_{{2}} \left( s \right)}{ \left( 1-\rho \right)  \left( s\,h \left(
s \right) -2\,\alpha\,\varepsilon \right) }}\spazio{0.8}\cr
&{}&{\it u_-} \left( s \right) =-{\frac { 2\sqrt{j(j+1)}u_{{2}} \left( s \right) }{s\,h \left( s \right)
}}-{\frac {2\rho\,u_{{3}} \left( s \right) }{ \left( 1-\rho \right) h \left( s \right) }}\spazio{0.8}\cr
&{}&{\it v_+} \left( s \right) ={\frac { 2\sqrt{j(j+1)}u_{{1}} \left( s \right) }{s\,h \left( s \right)
-2\,\alpha\,\varepsilon}}
-{\frac {2\rho\,s\,u_{{4}} \left( s \right) }{ \left( 1-\rho \right)  \left( s\,h \left( s \right)
-2\,\alpha\,\varepsilon \right) }}
\nonumber
\end{eqnarray}
We are eventually left with the system (\ref{system},\ref{even_coeff}).
By the final observation of Appendix B, the odd system and the corresponding algebraic relations are obtained
by letting 
$$\frac2{1-\rho} \rightarrow -\frac{2\rho}{1-\rho}\qquad \mathrm{and} \qquad 
\frac{2\rho}{1-\rho} \rightarrow -\frac2{1-\rho}.$$


\section*{APPENDIX D. eVP potentials} \label{appendixd}


We report here the potentials describing the vacuum polarization by electrons, firstly calculated by K\"allen and Sabry
and later on redetermined by other authors \cite{UKSK,Pachu}. For the sake of completeness we report
here the expressions we have used to calculate the muonic Hydrogen Lamb shift, given in the last of ref.
\cite{UKSK,Pachu}.
\begin{eqnarray}
V_\kappa = -\alpha\,\int\limits_0^1\,dv\,\rho_\kappa(v)\,\exp(-\lambda r)/r
\nonumber
\end{eqnarray}
where $\lambda=4m_e^2/\sqrt{1-v^2}$, $m_e$ being the electron mass and where the densities $\rho_\kappa(v)$ are defined
by
\begin{eqnarray}
&{}&\displaystyle \rho_{{1\phantom{1}}} \left( v \right) ={\frac {\alpha\,{v}^{2} \left( 1-1/3\,{v}^{2} \right) }{\pi \,
\left(
-{v}^{2}+1 \right) }}\spazio{1.4}\cr
&{}&\rho_{{11}} ( v ) =
-\,{\alpha}^{2}{v}^{2} \Bigl( 1-1/3\,{v}^{2} \Bigr)  \Bigl(
16-6\,{v}^{2}+\cr
&{}&\phantom{\rho_{{11}} ( v ) =}3\,v ( 3-{v}^{2} )\,\ln  \Bigl( {\frac {1-v}{1+v}} \Bigr)  \Bigr) 
\Bigl(9{\pi }^2 (1 -{v}^{2}) \Bigr)^{-1}\spazio{1.0}\cr
&{}& \rho_{{2}\phantom{1}} ( v ) =2{\alpha}^{2}v\,\Bigl(3\pi^2(1-v^2)  \Bigr)^{-1} 
\Bigl\{ ( 3-{v}^{2})  (1+ {v}^{2})\cr
&{}& \phantom{\rho_{{2}} ( v ) =}\Bigl[\,\, {\mathrm Li}_2 \Bigl( -{\frac {1-v}{1+v}} \Bigr) +2\,{\mathrm Li}_2 \Bigl(
{\frac {1-v}{1+v}} \Bigr)\cr
&{}&\phantom{\rho_{{2}} ( v ) =} +\ln  \Bigl(
{\frac {1+v}{1-v}} \Bigr)  \Bigl( \frac 32\,\ln  \Bigl( 1/2+1/2\,v \Bigr) -\ln(v)  \Bigr)\,\Bigr]  \cr
&{}& \phantom{\rho_{{2}} ( v ) =}+\ln 
\Bigl( {\frac {1+v}{1-v}} \Bigr) \Bigl( {\frac {11}{16}}\, ( 3-{v}^{2})  ( 1+{v}^{2}) +\frac14\,{v}^{4} \Bigr) \cr
&{}&\phantom{\rho_{{2}} ( v ) =} +\frac 32\,v\, ( 3-{v}^{2}) \ln  \Bigl( \frac{1-{v}^{2}}4 \Bigr) 
-2\,v ( 3-{v}^{3})\ln(v) \cr
&{}&\phantom{\rho_{{2}} ( v ) =} +\frac 38\,v\, ( 5-3\,{v}^{2} )  \Bigr\}
\nonumber
\end{eqnarray}


\begin{thebibliography}{999}

\bibitem{Eides} M.I. Eides, H. Grotch, V.A. Shelyuto, ``\emph{Theory of Light Hydrogenic Bound States}'',
Springer Tracts in Modern Physics 222, (Springer Verlag, Berlin 2007).

\bibitem{Karshen} S.G. Karshenboim, {Phys. Rep.}, {\bf 422}, 1, (2005).

\bibitem{GS1} R. Giachetti, E. Sorace, {J. Phys. A}, {\bf 38}, 1345, (2005).

\bibitem{GS2} R. Giachetti, E. Sorace, {J. Phys. A} {\bf 39}, 15207, (2006).

\bibitem{Lucha} W. Lucha, F. Sch\"oberl, arXiv 1407.4624 (2014) and references therein.

\bibitem{GS3} R. Giachetti, E. Sorace, {Phys. Rev. D}, {\bf 87}, 034021, (2013).

\bibitem{Richard} J.M. Richard, arXiv:1205.4326v2, (2012).

\bibitem{Rafelski} J.D. Carroll, A.W. Thomas, J. Rafelski, G.A. Miller, {Phys. Rev.} A {\bf 84}, 012506, (2011).

\bibitem{RR} S.F. Radford, W.W. Repko, {Phys. Rev.} D {\bf 75}, 074031, (2007);
{Nucl. Phys.} {\bf A865}, 69 (2011); 
A.M. Badalian, B.L.G. Bakker, I.V. Danilkin, {Phys. At. Nucl.}  {\bf 74}, 631, (2011).

\bibitem{Landau} V.B. Berestetskii, E.M. Lifshitz, L.P. Pitaevskii,\textit{ Quantum Electrodynamics} (Pergamon Press,
1982).

\bibitem{Karshenboim} S.G. Karshenboim, V.G. Ivanov. Eur. Phys. J. {D 19}, 13, (2002).

\bibitem{Jungmann} K-P. Jungmann, in ``\emph{The Hydrogen atom}'', LNP 570, 81, (2001) .

\bibitem{Pachu} K. Pachucki, {Phys. Rev. A} {\bf 53}, 2092, (1996). 

\bibitem{Martynenko}  E.N. Elekina, A.P. Martynenko, {Phys. Atom. Nucl.} {\bf 73}, 1828, (2010) and references therein
to all the previous papers on the muonic atom.

\bibitem{UKSK} E.A. Uheling, {Phys. Rev.} {\bf 48}, 55, (1935); G. K\"allen, A. Sabry, {K. Dan. Vidensk. Selsk.
Mat. Fys. Medd.},  {\bf 29}, No. 17, (1955); S.G. Karshenboim {Phys. Rev.D} {\bf 88}, 125019, (2013).

\end{thebibliography}
\end{document}